\documentclass[aps,prl,showpacs,twocolumn,superscriptaddress]{revtex4-1}
\pdfoutput=1
\usepackage{amsmath,amsthm,amssymb}
\usepackage{array}
\usepackage{graphicx}
\usepackage{fancyhdr}
\usepackage{color}
\usepackage{extarrows}
\usepackage{enumerate}
\usepackage{epstopdf}
\usepackage[colorlinks,
            linkcolor=black,
            citecolor=black,
            urlcolor=blue
            ]{hyperref}
\usepackage{natbib}

\newtheorem{definition}{Definition}

\newtheorem{example}{Example}

\begin{document}

\title{Determine dynamical behaviors by the Lyapunov function in competitive Lotka-Volterra systems}

\author{Ying Tang}
\email{Corresponding author. Email: james23@sjtu.edu.cn}
%\affiliation{Shanghai Center for Systems Biomedicine and Department of Physics\\Shanghai Jiao Tong University, Shanghai, 200240, China}
\affiliation{ZhiYuan college, Shanghai Jiao Tong University, China}
\affiliation{Shanghai Center for Systems Biomedicine, Shanghai Jiao Tong University, China}
\author{Ruoshi Yuan}
\author{Yian Ma}
\affiliation{Department of Computer Science and Engineering, Shanghai Jiao Tong University, China}
%Key Laboratory of Systems Biomedicine of Ministry of Education.

\date{\today}

\begin{abstract}
Global dynamical behaviors of the competitive Lotka-Volterra
system even in $3$-dimension are not fully understood. The Lyapunov function can provide us such knowledge once it is
constructed. In this paper, we construct explicitly the Lyapunov function in three examples of the competitive Lotka-Volterra system for the whole state space: $(1)$ the general $2$-dimensional case; $(2)$ a $3$-dimensional model; $(3)$ the model of May-Leonard. The dynamics of these examples include bistable case and cyclical behavior. The first two examples are the generalized gradient system defined in the Appendixes, while the model of May-Leonard is not. Our method is helpful to understand the limit cycle problems in general $3$-dimensional case.

\end{abstract}
\maketitle

\section{\textrm{I}. Introduction}
\label{section 1}
Lotka-Volterra system is one of the most fundamental models describing the interaction of $n$ species in mathematical ecology \cite{may2001stability}, physics and economics \cite{hofbauer1998evolutionary}. It is given by the following ordinary differential equations:
\begin{definition}[Lotka-Volterra system]
\begin{align}
\label{def:lotka}
\dot{x}_{i}=x_{i}\left(b_{i}-\sum^{n}_{j=1}a_{ij}x_{j}\right), i=1,\dots,n,
\end{align}
where each $x_{i} (i=1,\cdots,n)$ represents the population of one species and $b_{i}, a_{ij} (i=1, \cdots, n; j=1, \cdots, n)$ are constants depending on the environment. The state space of the system (\ref{def:lotka}) is represented by the non-negative vectors $
R^{n}_{+}=\{(x_{1}, \cdots, x_{n})\in R^{n}| x_{i}\geq 0,
i=1,\cdots,n\}$. When $b_{i}>0, a_{ij}>0 (i=1, \cdots, n)$, it is the competitive Lotka-Volterra system.
\end{definition}

Due to the nonlinear attributes of the competitive Lotka-Volterra system, its dynamics can be complex when $n\geq3$, such as cyclical behavior \cite{may1975nonlinear} and chaotic behavior \cite{wang2010bifurcations}. M. Hirsch has proved any trajectory of a $n$-dimensional competitive Lotka-Volterra system will converge to an invariant surface $\Sigma$, homeomorphic to $(n-1)$-dimensional unit simplex $\Delta={x_{i}: x_{i}\geq0, \sum_{i=1}^{n-1}x_{i}=1}$.

In $3$-dimensional case, following M. Hirsch's general result, M. L. Zeeman identified $33$ stable equivalence classes, of which only classes $26-31$ can have limit cycles. Then in \cite{xiao2000limit}, D. Xiao and W. Li proved the number of limit cycles is finite without a heteroclinic polycycle. In \cite{hofbauer1994multiple}, J. Hofbauer and J. So conjectured the number of limit cycles is at most two. Nevertheless, three limit cycles were constructed numerically in \cite{lu2003three,gyllenberg20063d} and four in \cite{gyllenberg2009four,wang2011limit}. M. L. Zeeman also tried to deduce global dynamics by the edges of the carrying simplex \cite{zeeman2002classification,zeeman2002n}. Till now, however, the question of how many limit cycles can appears in M. L. Zeeman's six classes $26-31$ remains open.

To analyze global dynamics, M. Planck studied the Lotka-Volterra system by hamiltonian theory, however, in limited parameter region \cite{plank1995hamiltonian}. The split Lyapunov function has been used \cite{zeeman2003local}, but it is not monotone along all the trajectories in the state space, hence constructed locally. Besides, the classical Lyapunov function has also been constructed for $n$-dimensional case in the parameter region with one global stable equilibrium \cite{y1996global,goh1977global}. As there is no general way of constructing the Lyapunov function \cite{strogatz2000nonlinear}, this theory has not been explored more to study the competitive Lotka-Volterra system.

In this paper, based on the framework of general dynamics recently proposed in \cite{ao2004potential,Yuan2010}, we construct the Lyapunov function in three examples of the competitive Lotka-Volterra system. This Lyapunov function is monotone along trajectories in the state space, thus can demonstrate global dynamics. The first example is the general $2$-dimensional case \cite{y1996global}. The second is a $3$-dimensional system given by \cite{y1996global} and the construction method is the same with the first's as both are the generalized gradient system defined in the Appendixes. The third example is the classical May-Leonard $3$-dimensional system \cite{may1975nonlinear}. As it is not the generalized gradient system, we provide there a different construction method.

This paper is organized as follows. In Sec. \textrm{II}\ref{section 2}, we uniformly construct the Lyapunov function for the $2$-dimensional model, and analyze its dynamics in the state space. In Sec. \textrm{III}\ref{section 3}, we study the $3$-dimensional competitive Lotka-Volterra system given by \cite{y1996global}. In Sec. \textrm{IV}\ref{section 4}, we study the model of May-Leonard. In Sec. \textrm{V}\ref{section 5}, we summarize our work. In the Appendixes, we introduce briefly our construction framework, discuss the generalized gradient system, and then give detailed calculation on other dynamical parts in our framework of the three examples.

\section{\textrm{II}. The general $2$-dimensional competitive Lotka-Volterra system}
\label{section 2}
\begin{example}
\label{ex:one}
The general $2$-dimensional competitive Lotka-Volterra system is given by:
\begin{equation}
\label{example1}
    \left\{
        \begin{array}{c}
            \dot{x}_{1}=x_{1}(b_{1}-x_{1}-\alpha x_{2})\\
            \dot{x}_{2}=x_{2}(b_{2}-\beta x_{1}-x_{2})
        \end{array}
    \right.~,
\end{equation}
where $b_{1}$, $b_{2}$, $\alpha$, $\beta$ are non-negative constants \cite{y1996global}. By setting $\dot{x}_{1}=\dot{x}_{2}=0$, four non-negative equilibriums are derived: (1) a positive one $E_{++}=\left(b_{1}-\alpha b_{2},b_{2}-\beta b_{1}\right)/(1-\alpha \beta)$ existing when $\alpha<b_{1}/b_{2}$, $\beta<b_{2}/b_{1}$ or $\alpha>b_{1}/b_{2}$, $\beta>b_{2}/b_{1}$; (2) $ E_{+0}=(b_{1},0)$; (3) $E_{0+}=(0,b_{2})$; (4) $E_{00}=(0,0)$. Here the subscript $+$ denotes the population of the species is positive and the subscript $0$ means the species dies out.
\end{example}

\subsection{A. Construction of the Lyapunov function}
\label{section 2.1}
Now we introduce our method to construct the Lyapunov function of the system. This method is based on the framework in \cite{ao2004potential,Yuan2010}. The idea is as follows. Assume there is a Lyapunov function $\phi$ and its partial derivative is given by
\begin{align*}
&\left(\begin{array}{c}
\frac{\partial\phi}{\partial x_{1}}\\
\frac{\partial\phi}{\partial x_{2}}
\end{array}\right)
\doteq-\left(\begin{array}{cc}
A_{11}(x_{1},x_{2})&A_{12}(x_{1},x_{2})\\
A_{21}(x_{1},x_{2})&A_{22}(x_{1},x_{2})
\end{array}\right)
\left(\begin{array}{c}
\dot{x}_{1}\\
\dot{x}_{2}
\end{array}\right)~,
\end{align*}
where $A_{11}(x_{1},x_{2})$, $A_{12}(x_{1},x_{2})$, $A_{21}(x_{1},x_{2})$, $A_{22}(x_{1},x_{2})$ are undetermined coefficients. Our aim is to choose proper coefficients so that: $(1) \nabla\times\nabla\phi=0$; $(2) \dot{\phi}\leq0$, i.e., Lie derivative of $\phi$ decreasing along trajectories.

We discover that $A_{11}(x_{1},x_{2})=\beta/x_{1}$, $A_{12}(x_{1},x_{2})=0$, $A_{21}(x_{1},x_{2})=0$, $A_{22}(x_{1},x_{2})=\alpha/x_{2}$ is a proper setting. Thus we get
\begin{align}
\label{partial}
\left\{
    \begin{array}{c}
    \frac{\partial\phi}{\partial x_{1}}=-\beta(b_{1}-x_{1}-\alpha x_{2})\\
    \frac{\partial\phi}{\partial x_{2}}=-\alpha(b_{2}-\beta x_{1}-x_{2})
    \end{array}~.
\right.
\end{align}
With direct calculation, $\nabla\times\nabla\phi=0$ and
\begin{align*}
\dot{\phi}&=\frac{\partial\phi}{\partial x_{1}}\dot{x}_{1}+\frac{\partial\phi}{\partial x_{2}}\dot{x}_{2}
\\&=-\beta x_{1}(b_{1}-x_{1}-\alpha x_{2})^{2}-\alpha x_{2}(b_{2}-\beta x_{1}-x_{2})^{2}\leq 0~,
\end{align*}
as $x_{1}$ and $x_{2}$ are all non-negative population species and $\beta$ and $\alpha$ are all non-negative constants. $\dot{\phi}(\mathbf{x})=0$ happens only at $\mathbf{x}\in\cup_{\mathbf{s}\in \mathbb{R}_{+}^2}\omega(\mathbf{s})$, where $\omega(\mathbf{s})$ denotes the $\omega$-limit set \cite{hirsch1974differential}. Thus, we can get a Lyapunov function by integrating the Eq. (\ref{partial}):
\begin{equation}
\phi=\frac{\beta}{2}x_{1}^{2}+\frac{\alpha}{2}x_{2}^{2}-\beta b_{1}x_{1}-\alpha b_{2}x_{2}+\alpha\beta x_{1}x_{2}~.
\end{equation}

Here we mention that the choice on the coefficients $A_{11}(x_{1},x_{2})$, $A_{12}(x_{1},x_{2})$, $A_{21}(x_{1},x_{2})$, $A_{22}(x_{1},x_{2})$ is not unique. Our choice is straightforward and meets the requirements.

\subsection{B. Analysis on dynamics in the state space}
\label{section 2.2}
\begin{figure}
%\setcaptionwidth{0.8\textwidth}
\centering
\includegraphics[width=0.5\textwidth]{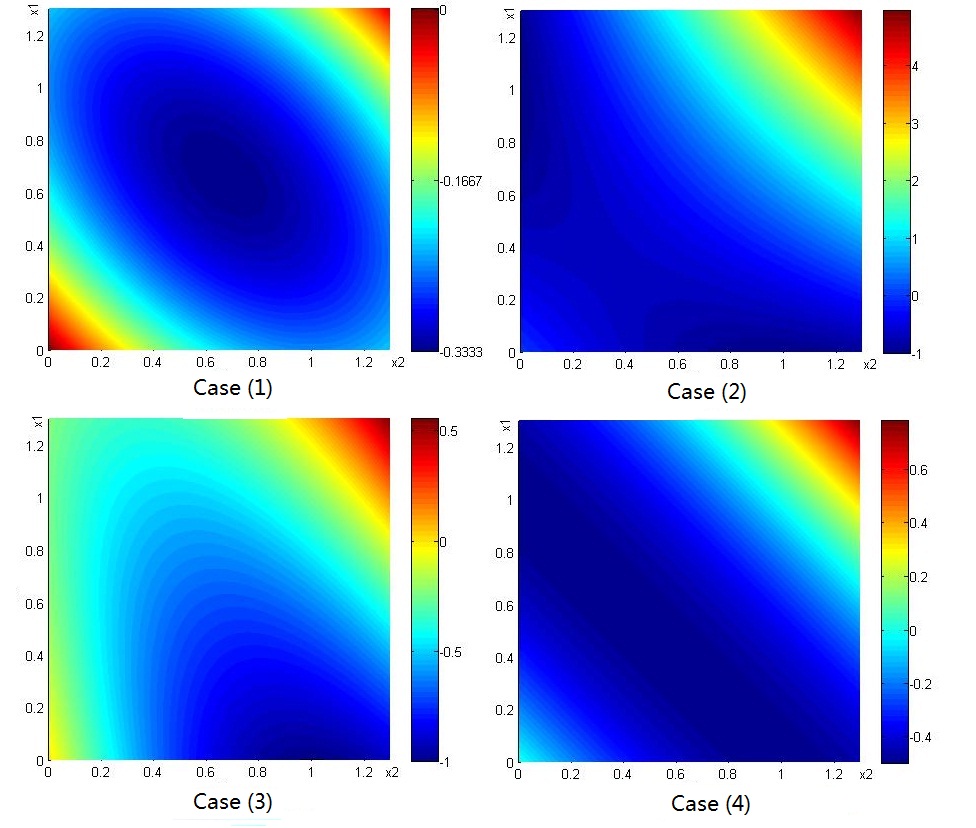}
\caption{The energy landscape of example (\ref{ex:one}) for the various cases: (1) $\alpha=\beta=\frac{1}{2}$, $b_{1}=b_{2}=1$: stable coexistence case; (2) $\alpha=\beta=2$, $b_{1}=b_{2}=1$: bistable case; (3) $\alpha=\frac{1}{2}$, $\beta=2$, $b_{1}=b_{2}=1$: one survival case; (4) $\alpha=\beta=b_{1}=b_{2}=1$: degenerate case.}
\label{fig:Global Lyapunov 1}
\end{figure}

In this subsection, we give a classified discussion on dynamics in each parameter region by the Hessian matrix of the Lyapunov function at $E_{++}$. As $\frac{\partial^{2}\phi}{\partial x_{1}^{2}}=\beta$, $\frac{\partial^{2}\phi}{\partial x_{2}^{2}}=\alpha$, $\frac{\partial^{2}\phi}{\partial x_{1}\partial x_{2}}=\frac{\partial^{2}\phi}{\partial x_{2}\partial x_{1}}=\alpha \beta$, we find the determinant of Hessian matrix:
\begin{align}
    \Delta\doteq\frac{\partial^{2}\phi}{\partial x_{1}^{2}} \frac{\partial^{2}\phi}{\partial x_{2}^{2}}-\left(\frac{\partial^{2}\phi}{\partial x_{1}\partial x_{2}}\right)^{2}=\alpha \beta(1-\alpha \beta)~.
\end{align}
Thus, the type of dynamics can be classified into four cases:

\begin{enumerate}[(1)]
  \item
  \label{case 1}
   Stable coexistence case: $\alpha<b_{1}/b_{2}$, $\beta<b_{2}/b_{1}$.

    $\Delta>0$ and $\frac{\partial^{2}\phi}{\partial x_{1}^{2}}>0$ indicate $E_{++}$ is a globally stable equilibrium with the minimum energy value.

  \item
  \label{case 2}
   Bistable case: $\alpha>b_{1}/b_{2}$, $\beta>b_{2}/b_{1}$.

  $\Delta<0$ indicates $E_{++}$ is a saddle point. As the system (\ref{example1}) is bounded in the first quadrant, it has two stable equilibriums $E_{+0}$ and $E_{0+}$ on the boundary.

  \item
  \label{case 3}
  One survival case: $\alpha<b_{1}/b_{2}$, $\beta>b_{2}/b_{1}$ or $\alpha>b_{1}/b_{2}$, $\beta<b_{2}/b_{1}$.

  It has one globally stable equilibrium on an axis of coordinate, $E_{+0}$ appears when $\alpha<b_{1}/b_{2}$, $\beta>b_{2}/b_{1}$ or $E_{0+}$ appears when $\alpha>b_{1}/b_{2}$, $\beta<b_{2}/b_{1}$. We just show the case where the species $x_{1}$ survives in Fig.~\ref{fig:Global Lyapunov 1}, i.e., when $\alpha<b_{1}/b_{2}$, $\beta>b_{2}/b_{1}$. The case where the species $x_{2}$ survives can be shown similarly.

  \item
  \label{case 4}
  Degenerate case: $\alpha=b_{1}/b_{2}$, $\beta=b_{2}/b_{1}$.

  The Lyapunov function has the minimum value along the line:\\ $\sqrt{b_{1}b_{2}}-\sqrt{b_{2}/b_{1}}x_{1}-\sqrt{b_{1}/b_{2}}x_{2}=0$ as in this case
  \begin{align}
  \phi=\frac{1}{2}\left(\sqrt{b_{1}b_{2}}-\sqrt{b_{2}/b_{1}}x_{1}-\sqrt{b_{1}/b_{2}}x_{2}\right)^{2}-b_{1}b_{2}~.
  \end{align}
  Each trajectory will converge to one of the points on the line, depending on the initial value.
\end{enumerate}

Four remarks are made here:
\begin{itemize}
  \item
  Our result on dynamics of the system is consistent with the  stability analysis near equilibriums in \cite{y1996global}. Additionally, our Lyapunov function is constructed uniformly for the whole parameter space and thus can provide dynamics for any perturbation on the parameters. Therefore, the criterion on the classification on the type of dynamics due to parameters changing can be based on the Lyapunov function: $(1)$ when $E_{++}$ exists, the Hessian matrix of the Lyapunov function at $E_{++}$ can show its type of stability and we have the previous two cases; $(2)$ when $E_{++}$ does not exist, we have the case (\ref{case 3}); $(3)$ the remaining one is the case (\ref{case 4}).
  \item
  Visualization with the energy landscape of the Lyapunov function (Fig.~\ref{fig:Global Lyapunov 1}) provides a clear observation on dynamics in each case above: Fig.~\ref{fig:Global Lyapunov 1}'s case (1) to case (4) relate to these cases (\ref{case 1}) to (\ref{case 4}) discussed above respectively. Saddle-node bifurcation can be observed from this energy landscape. The bifurcation happens from the case (\ref{case 1}) to the case (\ref{case 2}) and the degenerate case (\ref{case 4}) has the minimum energy value on a line.

  \item
  We show here the dynamics of the system can directly be determined by the Lyapunov function alone.
  In \cite{zeeman1993hopf}, M. L. Zeeman used the Lyapunov function to prove that the stable nullcline classes coincide with the stable topological classes in this system.

  \item
  In the Appendixes, we will define the generalized gradient system, as a natural generalization to typical gradient system. We thus find this $2$-dimensional system meet the definition. Furthermore, the  construction method can be applied to the generalized gradient system in $n$-dimension, and we will show a $3$-dimensional example in next section.
\end{itemize}

\section{\textrm{III}. A $3$-dimensional model}
\label{section 3}

Following the definition of the generalized gradient system in the Appendixes, we find that a $3$-dimensional competitive Lotka-Volterra system given by \cite{y1996global} is another one and thus we construct the Lyapunov function by the same method as the last section's.
\begin{example}
\label{ex:three}
\begin{align}
\label{example3}
\left\{
     \begin{array}{l}
        \dot{x}_{1}=x_{1}\left(1-x_{1}-\alpha x_{2}\right) \\
        \dot{x}_{2}=x_{2}\left(1-\beta x_{1}-x_{2}-\beta x_{3}\right) \\
        \dot{x}_{3}=x_{3}\left(1-\alpha x_{2}-x_{3}\right)
      \end{array}
   \right.~,
\end{align}
where $\alpha, \beta$ are non-negative coefficients. By setting $\dot{x}_{1}=\dot{x}_{2}=\dot{x}_{3}=0$, non-negative equilibriums are derived: (1) a positive one $E_{+++}=\left(1-\alpha,1-2\beta,1-\alpha\right)/(1-2\alpha\beta)$ existing when $(1-\alpha)/(1-2\alpha\beta)>0$ and $(1-2\beta)/(1-2\alpha\beta)>0$; (2) $ E_{+0+}=(1,0,1)$; (3) $E_{0+0}=(0,1,0)$; (4) $E_{000}=(0,0,0)$.
\end{example}

\subsection{A. Construction of the Lyapunov function}
With the similar method used in the general $2$-dimensional competitive Lotka-Volterra system, we choose the corresponding undetermined matrix to be
\begin{align*}
\left(\begin{array}{ccc}
\beta/x_{1} & 0 & 0\\
0 & \alpha/x_{2} & 0\\
0 & 0 & \beta/x_{3}
\end{array}\right),
\end{align*}
then
\begin{align}
\label{partial2}
\left\{
     \begin{array}{l}
        \frac{\partial\phi}{\partial x_{1}}=-\beta\left(1-x_{1}-\alpha x_{2}\right) \\
        \frac{\partial\phi}{\partial x_{2}}=-\alpha\left(1-\beta x_{1}-x_{2}-\beta x_{3}\right) \\
        \frac{\partial\phi}{\partial x_{3}}=-\beta\left(1-\alpha x_{2}-x_{3}\right)
      \end{array}
   \right.~.
\end{align}

Since $\nabla\times\nabla\phi=0$ and the Lie derivative of $\phi$ is
\begin{align*}
\dot{\phi}=&-\beta x_{1}\left(1-x_{1}-\alpha x_{2}\right)^{2}-\alpha x_{2}\left(1-\beta x_{1}-x_{2}-\beta x_{3}\right)^{2}\\
&-\beta x_{3}\left(1-\alpha x_{2}-x_{3}\right)^{2}\leq 0~,
\end{align*}
as $x_{1}, x_{2}$ and $x_{3}$ are all non-negative population species and $\beta$ and $\alpha$ are all non-negative constants. $\dot{\phi}(\mathbf{x})=0$ happens only at $\mathbf{x}\in\cup_{\mathbf{s}\in \mathbb{R}_{+}^3}\omega(\mathbf{s})$. We can construct a Lyapunov function by integrating the Eq. (\ref{partial2}).
\begin{equation}
\phi=\frac{\beta}{2}(x_{1}^{2}+x_{3}^{2})+\frac{\alpha}{2}x_{2}^{2}+\alpha\beta(x_{1}x_{2}+x_{2}x_{3})-\beta(x_{1}+x_{3})-\alpha x_{2}~.
\end{equation}

\subsection{B. Analysis on dynamics in the state space}
With the Lyapunov function constructed globally on $R^{3}_{+}$, first the classified stability analysis near equilibriums by \cite{y1996global} can be unified now. Second, when $\alpha=1$ and $\beta=1/2$,
\begin{align}
\phi=\frac{1}{4}\left[(x_{1}+x_{2}-1)^{2}+(x_{2}+x_{3}-1)^{2}-2\right]
\end{align}
 indicates the degenerate case of the system has the minimum energy on the intersection of the surfaces $x_{1}+x_{2}-1=0$  and $x_{2}+x_{3}-1=0$. Third, as it is a generalized gradient system without a trajectory contouring along the energy landscape of the Lyapunov function, the system (\ref{example3}) does not have a limit cycle \cite{Yuan2010}.

\section{\textrm{IV}. The model of May-Leonard}
\label{section 4}
In \cite{may1975nonlinear}, R. May and W. Leonard studied a $3$-dimensional competitive Lotka-Volterra system.
\begin{example}
\label{example4}
\begin{align}
\left\{
     \begin{array}{l}
        \dot{x}_{1}=x_{1}\left(1-x_{1}-\alpha x_{2}-\beta x_{3}\right) \\
        \dot{x}_{2}=x_{2}\left(1-\beta x_{1}-x_{2}-\alpha x_{3}\right) \\
        \dot{x}_{3}=x_{3}\left(1-\alpha x_{1}-\beta x_{2}-x_{3}\right)
      \end{array}
   \right.~,
\end{align}
where $\alpha,\beta$ are non-negative coefficients. The possible equilibriums contain: $(1) (0,0,0)$; $(2)$ three single-population survive $(1,0,0),(0,1,0),(0,0,1)$; $(3)$ three two-population solutions of the form $\frac{(1-\alpha,1-\beta,0)}{1-\alpha\beta}$; $(4)$ and three-population survive $\frac{(1,1,1)}{1+\alpha+\beta}$.
\end{example}

\subsection{A. Construction of the Lyapunov function}
We find that this model is not the generalized gradient system, and thus the construction method in Sec. \textrm{II}\ref{section 2} can not be applied here. So we will give another method to construct the Lyapunov function in this section.

For convenience, let us introduce some new variables:
$\gamma=\alpha+\beta-2$, $P=x_{1}x_{2}x_{3}$ and $O=x_{1}+x_{2}+x_{3}$. Then
\begin{align*}
\dot{P}&=\dot{x}_{1}x_{2}x_{3}+x_{1}\dot{x}_{2}x_{3}+x_{1}x_{2}\dot{x}_{3}
\\&=P\left[3-(1+\alpha+\beta)O\right]
\\&=P\left[3-(3+\gamma)O\right]
\\&=P\left[3(1-O)-\gamma O\right],
\end{align*}
and
\begin{align*}
\dot{O}&=\dot{x}_{1}+\dot{x}_{2}+\dot{x}_{3}
\\&=O-\left[x_{1}^{2}+x_{2}^{2}+x_{3}^{2}+(\alpha+\beta)(x_{1}x_{2}+x_{2}x_{3}+x_{3}x_{1})\right]
\\&=O(1-O)-\gamma(x_{1}x_{2}+x_{2}x_{3}+x_{3}x_{1}),
\end{align*}
where the $\dot{P}$ and $\dot{O}$ denotes the Lie derivative of $P$ and $O$ respectively. Next, we construct the Lyapunov function in two different parameter regions: $(1) \gamma=0$; $(2) \gamma\neq0$.

\begin{enumerate}
\item
    When $\gamma=0$:

    Noting that $\dot{P}=3P(1-O)$ and $\dot{O}=O(1-O)$, thus
    \begin{align}
    -\frac{\dot{P}}{P}+3\dot{O}=-3(1-O)^{2}\leq0.
    \end{align}
    So if we can a function whose Lie derivative is $-\dot{P}/P+3\dot{O}$, then it is a Lyapunov function. This can be done by simply integrate $-\dot{P}/P+3\dot{O}$ and we get a Lyapunov function
    \begin{align}
    \phi&=3O-\ln P
    \\&=3(x_{1}+x_{2}+x_{3})-\ln(x_{1}x_{2}x_{3}).
    \end{align}
    $\dot{\phi}=0$ only when $O-1=0$, i.e., all the trajectories converge to the plane $O=1$.

    \item
    When $\gamma\neq0$:

    Noting that $\dot{O}=O(1-O)-\gamma(x_{1}x_{2}+x_{2}x_{3}+x_{3}x_{1})$ and $\dot{P}=P\left[3(1-O)-\gamma O\right]$, thus
    \begin{align*}
    \quad\quad&\gamma\left[\dot{P}O-3\dot{O}P\right]
    \\&=\gamma\Big[3PO(1-O)-\gamma PO^{2}-3PO(1-O)
    \\&\quad+3\gamma P(x_{1}x_{2}+x_{2}x_{3}+x_{3}x_{1})\Big]
    \\&=-\gamma^{2}P\Big[O^{2}-3(x_{1}x_{2}+x_{2}x_{3}+x_{3}x_{1})\Big]
    \\&=-\gamma^{2}P\Big[x_{1}^{2}+x_{2}^{2}+x_{3}^{2}-(x_{1}x_{2}+x_{2}x_{3}+x_{3}x_{1})\Big]
    \\&=-\frac{\gamma^{2}P}{2}\Big[(x_{1}-x_{2})^{2}+(x_{2}-x_{3})^{2}+(x_{3}-x_{2})^{2}\Big]\leq0.
    \end{align*}
    So we need to find a function whose Lie derivative is $\gamma\left[\dot{P}O-3\dot{O}P\right]$. We notice that we can not integrate it directly, however, we find out that the function \begin{align}
    \phi=\gamma\frac{P}{O^{3}}
    \end{align}
    has the Lie derivative
    \begin{align}
    \dot{\phi}&=\gamma\frac{\dot{P}O-3\dot{O}P}{O^{4}}\leq0.
    \end{align}

    As a result, we get a Lyapunov function.
    $\dot{\phi}=0$ can happen at a point on the line $x_{1}=x_{2}=x_{3}$ or in the set ${(x_{1},x_{2},x_{3})|P=0}$.
\end{enumerate}

\subsection{B. Analysis on dynamics in the state space}

With the Lyapunov function constructed in the last subsection, we discuss dynamics in classified parameter space of the system here.
\begin{enumerate}
    \item $\gamma=0$:

    As $\dot{\phi}=0$ only on the plane $O=1$, all the trajectories will converge to this plane. Besides, on this plane, the value of the Lyapunov function $\phi=3+\ln P$ will be a constant. Thus for each trajectory on the plane, the value of $P$ will be a constant. This means that the limit set for any initial point will be the intersection of the plane $O=1$ and the hyperboloid that $P$ equals to a constant, which is a cycle on the plane.

    \item $\gamma<0$:

    As $\phi=\gamma\frac{P}{O^{3}}$ is non-positive now, the minimum value of $\phi$ will not be zero if its initial value is not. Thus, in order to minimize the value of the Lyapunov function, all the trajectory will converge to one point on the line $x_{1}=x_{2}=x_{3}$. So this case has a global stable equilibrium.

    \item $\gamma>0$:

    As $\phi=\gamma\frac{P}{O^{3}}$ is non-negative now, the minimum value of $\phi$ will be zero. That is, all the trajectory will converge to the set ${(x_{1},x_{2},x_{3})|P=0}$. In the neighborhood of $P=0$, the terms of order $x_{1}x_{2}$, etc., in $\dot{O}$ asymptotically make a negligible contribution \cite{may1975nonlinear}. Thus $\dot{O}=O(1-O)$ leads to $O\rightarrow1$ in the end. Finally, the limit set in this case is the set ${(x_{1},x_{2},x_{3})|P=0, O=1}$.
\end{enumerate}

\begin{figure*}
\includegraphics[width=1\textwidth]{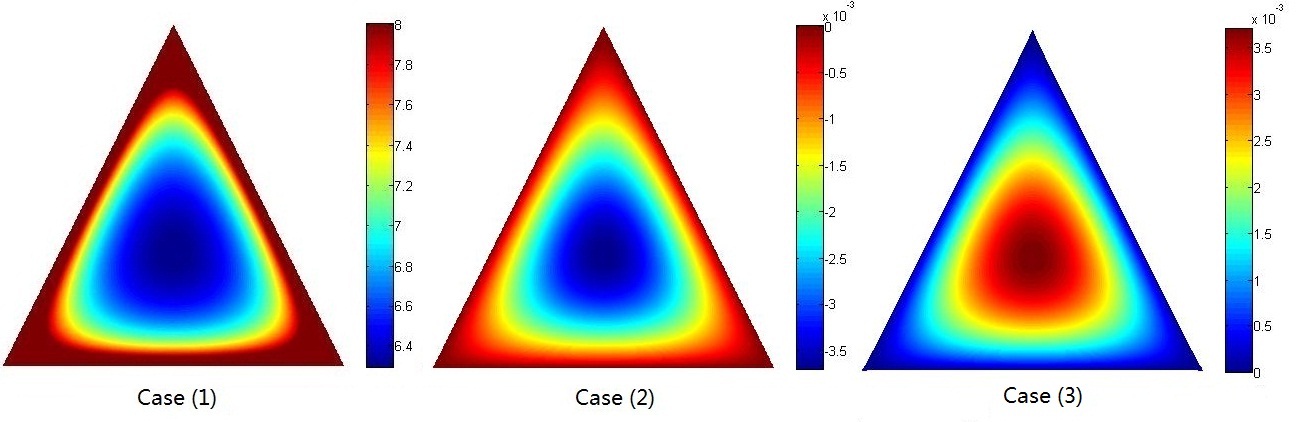}
\caption{The energy landscape of example (\ref{example4}) on the plane $O=1$: $(1)$ $\gamma=0$; $(2)$ $\gamma=-0.1<0$; $(3)$ $\gamma=0.1>0$.}
\label{fig:Global Lyapunov 2}
\end{figure*}

Three remarks are made here:
\begin{itemize}
  \item
  Our result is consistent with that in \cite{may1975nonlinear}. Furthermore, we give a full description on dynamics for the whole state space with the Lyapunov function. The energy landscape of the Lyapunov function on the plane $x_{1}+x_{2}+x_{3}$ (Fig.~\ref{fig:Global Lyapunov 2}) gives a direct observation on dynamics: (1)when $\gamma=0$, Fig.~\ref{fig:Global Lyapunov 2}'s case $(1)$ shows the system has hamiltonian structure; (2)when $\gamma>0$, Fig.~\ref{fig:Global Lyapunov 2}'s case $(2)$ shows the system has a global stable equilibrium; (3)when $\gamma<0$, Fig.~\ref{fig:Global Lyapunov 2}'s case $(3)$ shows the limit set of the system is ${(x_{1},x_{2},x_{3})|P=0, O=1}$.
  \item
  A similar analytic form of the Lyapunov function have been constructed when $\gamma>0$ in \cite{robinson2004introduction,hofbauer1981occurrence}. Compared with theirs, our construction is for the whole parameter space. Besides, we here provide a explicit method to find this Lyapunov function.
  \item
  In \cite{chi1998asymmetric}, C. W. Chi study the asymmetric May-Leonard system. Our construction method here may be able to be generalized to their system.
\end{itemize}

\section{\textrm{V}. Conclusion}
\label{section 5}
We have demonstrated that the Lyapunov function can be constructed in general $2$-dimensional and two $3$-dimensional competitive Lotka-Volterra systems. For each example, we have shown dynamics in the whole state space with the Lyapunov function. The $2$-dimensional case includes the bistable case and the model of May-Leonard has cycles as its limit set. Besides, in the Appendixes, we have defined the generalized gradient system and discussed its coherence and generality with the classical gradient system. Furthermore, we notice that the construction method used in the model of May-Leonard may be able to be generalized to the asymmetric May-Leonard system in \cite{chi1998asymmetric}. Thus the Lyapunov function can be helpful to solve the limit cycle problems in general $3$-dimensional case.

\section{Acknowledgment}
The authors would like to express their sincere gratitude for the helpful discussions with Ping Ao, Bo Yuan, Xinan Wang, Song Xu, Siyun Yang, Tianqi Chen and Jianghong Shi. This work was supported in part by the National 973 Projects No.~2010CB529200 and by the Natural Science Foundation of China No.~NFSC61073087 and No.~NFSC91029738.

\section*{Appendixes}
%\appendix
%\section{A}
%\setcounter{section}{1}
\label{appendix 1}
In the Appendixes, we first give the definition of the Lyapunov function and  the construction framework in the Appendix \textrm{I}\ref{section 6}. Then we introduce the generalized gradient system in the Appendix \textrm{II}\ref{section 7}. Finally in the Appendix \textrm{III}\ref{section_8}, we give detailed calculation on all the dynamical parts in our framework of the third example.

\section{Appendix \textrm{I}. The Lyapunov function}
\label{section 6}
\begin{definition}
A smooth dynamical system is given by
\begin{align}
\label{def:dynamical}
\dot{\mathbf{x}}=\mathbf{f}(\mathbf{x})\,,
\end{align}
where $\mathbf{x}=(x_{1},\cdots,x_{n})$ with $x_{1},\cdots,x_{n}$ the $n$ Cartesian coordinates of the state space, $\dot{\mathbf{x}}=d\mathbf{x}/dt$ and $\mathbf{f}:\mathbb{R}^n\mapsto\mathbb{R}^n$.
\end{definition}

The conventional Lyapunov function for a given system (\ref{def:dynamical}) is defined as:
\begin{definition}[Conventional Lyapunov Function]\cite{hirsch1974differential}
\label{def:Lyapunov}

Let $L: \mathcal{O}\to\mathbb{R}$ be a $C^1$ function, where $\mathcal{O}$ is an open set in $\mathbb{R}^n$. $L$ is a conventional Lyapunov function of the system (\ref{def:dynamical}) on $\mathcal{O}$ if

\begin{itemize}
\item
    \label{condition 1}
    for a specified equilibrium $\mathbf{x}^*$ in $\mathcal{O}$, $L(\mathbf{x}^*)=0$ and $L(\mathbf{x})>0$ when $\mathbf{x}\neq \mathbf{x}^*$;
\item $\dot{L}(\mathbf{x})=\frac{dL}{dt}|_\mathbf{x}\leqslant0$ for all $\mathbf{x}\in\mathcal{O}$.
\end{itemize}
\end{definition}

La Salle has extended the conventional Lyapunov function to include stable region by abandoning the positive definite requirement, but his generalization is too rough to lose stability information inside the stable region.

\begin{definition}[La Salle's Lyapunov Function]\cite{lasalle1976stability}
\label{def:LaSalle Lyapunov}

Let $L: \mathcal{O}\to\mathbb{R}$ be a $C^1$ function, where $\mathcal{O}$ is an open set in $\mathbb{R}^n$. $L$ is a La Salle's Lyapunov function of the system (\ref{def:dynamical}) on $\mathcal{O}$ if $\dot{L}(\mathbf{x})=\frac{dL}{dt}|_\mathbf{x}\leqslant0$ for all $\mathbf{x}\in\mathcal{O}$.
\end{definition}

Following the Lyapunov function used in \cite{ao2004potential,Yuan2010}, here we give a more precise definition on the Lyapunov function.

\begin{definition}[Lyapunov Function]
\label{def:cand_Lyapunov}

Let $\phi:\mathbb{R}^n\mapsto\mathbb{R}$ be a $C^1$ function. $\phi$ is a Lyapunov function of the system (\ref{def:dynamical}) if $\dot{\phi}(\mathbf{x})=\frac{d\phi}{dt}|_\mathbf{x}\leqslant0$ for all $\mathbf{x}\in \mathbb{R}^n$ and $\dot{\phi}(\mathbf{x})=0$ only when $\mathbf{x}$ belongs to the union of the $\omega$-limit sets $\cup_{\mathbf{s}\in \mathbb{R}^n}\omega(\mathbf{s})$.
\end{definition}

In the following, we will briefly introduce our construction framework. It is recently discovered during the study on the stability problem of a genetic switch \cite{zhu2004calculating,ao2004potential} and has been found wildly useful in biology \cite{ao2009global}.
The key result of the framework is a transformation from the $n$ dimensional system (\ref{def:dynamical}) to the vector differential equation (for simplicity, we only discuss the deterministic case in this paper, with noise strength being zero, general results with randomness can be found in \cite{ao2004potential}):
\begin{align}
\label{eq:CFE}
\left[S(\mathbf{x})+T(\mathbf{x})\right] \dot{\mathbf{x}}=-\nabla \phi (\mathbf{x})~.
\end{align}
Here $\phi$ is a scalar function, the Lyapunov function. $S$ is a semi-positive definite and symmetric matrix. $T$ is a antisymmetric matrix.

Symmetrically, if $(S+T)$ is nonsingular, the Eq. (\ref{eq:CFE}) can be rewritten as a reverse form
\begin{align}
\label{eq:SFE}
\dot{\mathbf{x}}=-\left[D(\mathbf{x})+Q(\mathbf{x})\right]\nabla \phi (\mathbf{x})\,,
\end{align}
where $D$ is a semi-positive definite and symmetric matrix, $Q$ is a antisymmetric matrix.

From a physical point of view, $S$ can be explained as a frictional force indicating dissipation of the potential energy, $T$ as a Lorentz force and $\phi$ as a potential of the system influenced by the two forces. Symbol $D$ denotes the diffusion matrix indicating the random driving force, therefore for deterministic systems, $D$ is free to choose.

We make four remarks here:
\begin{itemize}
   \item
   As $\dot{\phi}=\dot{\mathbf{x}}^{\tau}\nabla\phi=-\dot{\mathbf{x}}^{\tau}[S+T]\dot{\mathbf{x}}=-\dot{\mathbf{x}}^{\tau}S\dot{\mathbf{x}}\leq0$, where $\tau$ denotes transpose, $\phi$ in the Eq. (\ref{eq:CFE}) can be a Lyapunov function.
 \item
   In this decomposition of the dynamical system, $S$ can be considered as gradient part and $T$ as rotational part. When $S=0$, it is a conserved system with first integral. If $T$ is a scalar matrix at the same time, it is a Hamiltonian system where the trajectory would be a contour along the energy landscape of the Lyapunov function. When $T=0$, it is a generalized gradient system defined in the next section. Thus both $S$ and $T$ can provide dynamical information for a given system.

   \item
   If the explicit form of the Lyapunov function is not obtained yet for a given system, such as the third example, we can solve the Eq. (\ref{eq:SFE}) by choosing a suitable form of $D$ (or $S$) to make $\phi$ satisfy $\nabla\times\nabla\phi=0$ and $\dot{\phi}\leq0$.  Here $\nabla\times$ is a matrix generalization of the ``curl operator in 3 dimension": $\left(\nabla\times\dot{\mathbf{x}}\right)_{ij}\doteq\partial_{i}\dot{x}_{j}-\partial_{j}\dot{x}_{i}$. We have constructed the Lyapunov function by this method in the first two examples.
   \item
   If one already has a Lyapunov function for a given system, we can obtain other dynamical parts \cite{Yuan2010}:
   \begin{align}
   \label{S,T}
   S&=-\frac{\nabla \phi\cdot\mathbf{f}}{\mathbf{f}\cdot\mathbf{f}}I~,\\
   T&=-\frac{\nabla \phi\times \mathbf{f}}{\mathbf{f}\cdot\mathbf{f}}~.
   \end{align}
   The corresponding explicit expression of the diffusion matrix $D$ and the antisymmetric matrix $Q$ can be provided as well:
   \begin{align}
   D&=-\left[\frac{\mathbf{f}\cdot\mathbf{f}}{\nabla \phi \cdot \mathbf{f}}
   I+\frac{\left(\nabla \phi\times \mathbf{f}\right)^2}{\left(\nabla \phi \cdot \mathbf{f}\right)\left(\nabla \phi\cdot\nabla \phi\right)}\right]~, \\
   Q&=\frac{\nabla \phi\times \mathbf{f}}{\nabla \phi\cdot\nabla \phi}~.
   \end{align}
\end{itemize}

\section{Appendix \textrm{II}. Generalized gradient system}
\label{section 7}
\subsection{A. Definition}
\label{section 7.1}
\begin{definition}[Generalized Gradient System]
A generalized gradient system on $\mathbb{R}^{n}$ is a dynamical system of the form
\begin{align}
\label{def:Generalized Gradient System}
\dot{\mathbf{x}}=-D(\mathbf{x})\nabla \phi (\mathbf{x})~,
\end{align}
where $\phi:\mathbb{R}^n\mapsto\mathbb{R}$ is a continuous differentiable scalar function and $D(\mathbf{x})$ is a semi-positive definite and symmetric matrix.
\end{definition}

By definition, when $D$ is the product of a nonzero constant and the identity matrix, it degenerates to the classical gradient system in \cite{hirsch1974differential}. The potential gradient of the system (\ref{def:Generalized Gradient System}) is anisotropic, different from that of the gradient system. Such anisotropic system has been observed in real systems, like the Fourier's equation in \cite{de2009thermal}. $T=0$ is equivalent to $Q=0$ as
\begin{align*}
T(\mathbf{x})=\frac{1}{2}\left[(D(\mathbf{x})+Q(\mathbf{x}))^{-1}-(D(\mathbf{x})-Q(\mathbf{x}))^{-1}\right]~.
\end{align*}

\subsection{B. Matrices $S$, $T$, $D$ and $Q$ of the first two examples}

For the given system (\ref{example1}) in Sec. \textrm{II}\ref{section 2}, it is not a gradient system as the curl of the vector field $\nabla\times\dot{\mathbf{x}}=\alpha x_{1}-\beta x_{2}\neq0$. But we can calculate the matrices using the results obtained in Sec. \textrm{II}. A\ref{section 2.1}:
\begin{align*}
S=\left(\begin{array}{cc}
\beta/x_{1} & 0\\
0 & \alpha/x_{2}
\end{array}\right),\quad T=0,
\\ \quad D=\left(\begin{array}{cc}
x_{1}/\beta & 0\\
0 & x_{2}/\alpha
\end{array}\right), \quad Q=0.
\end{align*}
$D$ being semi-positive definite and symmetric and $T=0$ indicate that the system (\ref{example1}) is a generalized gradient system with zero rotational part, and trajectory will not contour along the energy landscape of the Lyapunov function. Besides, $S$  is singular only on the coordinate axis in this system. It means the dissipation is infinite and thus the trajectory will stay on the axis once reaching it and approach the equilibrium $E_{+0}=(b_{1},0)$ or $E_{0+}=(0,b_{2})$.

As for the system (\ref{example2}) in Sec. \textrm{III}\ref{section 3}, the matrices are
\begin{align*}
S=\left(\begin{array}{ccc}
\beta/x_{1} & 0 & 0\\
0 & \alpha/x_{2} & 0\\
0 & 0 &\beta/x_{3}
\end{array}\right),\quad T=0,
\\ \quad D=\left(\begin{array}{ccc}
x_{1}/\beta & 0 & 0\\
0 & x_{2}/\alpha & 0\\
0 & 0 & x_{3}/\beta
\end{array}\right), \quad Q=0.
\end{align*}

\subsection{C. Linear Cases}
\label{section 7.2}
A linear autonomous dynamical system is given by the following ordinary differential equations:
\begin{align}
\label{def:linear dynamical}
\dot{\mathbf{x}}=F\mathbf{x}~,
\end{align}
where $\mathbf{x}=(x_{1},\cdots,x_{n})$ with $x_{1},\cdots,x_{n}$ the $n$ Cartesian coordinates of the state space, $\dot{\mathbf{x}}=d\mathbf{x}/dt$ and $F$ is a constant matrix. To ensure the independence of all the state variables, we require the determinant of the $F$ matrix to be finite: $det(F)\neq0$.

To illustrate the coherence and generality of the generalized gradient system in the linear cases, we first mention that a linear system (\ref{def:linear dynamical}) is a gradient system $\dot{\mathbf{x}}=-\nabla\phi$ if and only if its $F$ matrix is symmetric:
\begin{itemize}
  \item A gradient system $\dot{\mathbf{x}}=-\nabla\phi$ has $\partial{\phi}/\partial{x_{i}}=-\Sigma^{n}_{j=1}F_{ij}x_{j}$, then $\nabla\times\nabla\phi=0$ leads to the $F$ matrix being symmetric;
  \item If a linear system (\ref{def:linear dynamical}) has a symmetric $F$ matrix, then by setting $\partial{\phi}/\partial{x_{i}}=-\Sigma^{n}_{j=1}F_{ij}x_{j}$, the solution of $\phi$ exists, and we can rewrite (\ref{def:linear dynamical}) as $\dot{\mathbf{x}}=-\nabla\phi$.
\end{itemize}

But a linear system (\ref{def:linear dynamical}) can be a generalized gradient system when the $F$ matrix is asymmetric. Such systems have nonzero curl, that is $\nabla\times\dot{\mathbf{x}}\neq0$. We give an example of $2$ dimensional linear generalized gradient system in the following.

\begin{example}
\label{ex:two}
This example is given by \cite{hirsch1974differential}:
\begin{align}
\label{example2}
&\left(\begin{array}{c}
\dot{x}_{1}\\
\dot{x}_{2}
\end{array}\right)
=\left(\begin{array}{cc}
0&3\\
1&-2
\end{array}\right)
\left(\begin{array}{c}
{x}_{1}\\
{x}_{2}
\end{array}\right)~.
\end{align}
\end{example}
We set a Lyapunov function to be $\phi=x_{2}^{2}-x_{1}x_{2}$ as its Lie derivative
\begin{align*}
\dot{\phi}=-3x_{2}^{2}-\left(x_{1}-2x_{2}\right)^{2}\leq0~.
\end{align*}
Then the system (\ref{example2}) can be rewritten as
\begin{align}
&\left(\begin{array}{c}
\dot{x}_{1}\\
\dot{x}_{2}
\end{array}\right)
=-\left(\begin{array}{cc}
3&0\\
0&1
\end{array}\right)
\left(\begin{array}{c}
\frac{\partial\phi}{\partial x_{1}}\\
\frac{\partial\phi}{\partial x_{2}}
\end{array}\right)~.
\end{align}
Therefore, the original system (\ref{example2}) is a generalized gradient system by definition, but not a gradient system as its $F$ matrix is asymmetric.

\section{\textrm{III}. Matrices $S$ and $T$ for the model of May-Leonard}
\label{section_8}
In this section, we calculate $S$ and $T$ by the Eq. (\ref{S,T}) for the model of May-Leonard system. In Sec. \textrm{V}\ref{section 5}, we have obtained the Lyapunov function:
\begin{enumerate}
\item
    When $\gamma=0$:
    \begin{align*}
    \phi&=3O-\ln P
    \\&=3(x_{1}+x_{2}+x_{3})-\ln(x_{1}x_{2}x_{3}).
    \end{align*}
\item
    When $\gamma\neq0$:
   \begin{align*}
    \phi&=\gamma\frac{P}{O^{3}}
    \\&=\gamma\frac{x_{1}x_{2}x_{3}}{(x_{1}+x_{2}+x_{3})^{3}}.
    \end{align*}
\end{enumerate}
Thus we do calculation separately for this two cases in the following.
\begin{enumerate}
\item
    When $\gamma=0$:

    As
    \begin{align}
    \left\{
     \begin{array}{l}
        \frac{\partial\phi}{\partial x_{1}}=\frac{3x_{1}-1}{x_{1}} \\
        \frac{\partial\phi}{\partial x_{2}}=\frac{3x_{2}-1}{x_{2}} \\
        \frac{\partial\phi}{\partial x_{3}}=\frac{3x_{3}-1}{x_{3}}
      \end{array}
    \right.~,
    \end{align}
    \begin{align}
   S&=-\frac{\nabla
   \phi\cdot\mathbf{f}}{\mathbf{f}\cdot\mathbf{f}}I
   \notag\\&=\frac{3[1-(x_{1}+x_{2}+x_{3})]^{2}}{\dot{x}_{1}^{2}+\dot{x}_{2}^{2}+\dot{x}_{3}^{2}}I~.
   \end{align}
   Notice on the plane $x_{1}+x_{2}+x_{3}=1$, $S$ is zero matrix, thus on the plane the system is conserved when $\gamma=0$.

   As for $T$:
   \begin{align}
   T&=-\frac{\nabla \phi\times \mathbf{f}}{\mathbf{f}\cdot\mathbf{f}}
   \notag\\&=-\frac{\Big(\frac{3x_{i}-1}{x_{i}}\dot{x}_{j}-\frac{3x_{j}-1}{x_{j}}\dot{x}_{i}\Big)_{3\times3}}{\dot{x}_{1}^{2}+\dot{x}_{2}^{2}+\dot{x}_{3}^{2}}~.
   \end{align}

   Since $T$ is antisymmetric, we just need to calculate the elements  $T_{12}$, $T_{13}$ and $T_{23}$ of the above $3\times3$ matrix. As we have proved all the trajectory will converge to the plane $x_{1}+x_{2}+x_{3}=1$, we calculate the elements on the plane below. We first calculate $T_{12}$.
   \begin{align}
    T_{12}&=\frac{3x_{1}-1}{x_{1}}\dot{x}_{2}-\frac{3x_{2}-1}{x_{2}}\dot{x}_{1}
    \notag\\&=\frac{3x_{1}-1}{x_{1}}(1-\beta x_{1}-x_{2}-\alpha x_{3})x_{2}
    \notag\\&\quad-\frac{3x_{2}-1}{x_{2}}(1-x_{1}-\alpha x_{2}-\beta x_{3})x_{1}~.
    \end{align}
    We notice that $1-\alpha=-(1-\beta)=\frac{\beta-\alpha}{2}$ in the case of $\gamma=0$. Thus we have
    \begin{align}
    T_{12}&=\frac{\beta-\alpha}{2}\Big[\frac{x_{2}}{x_{1}}[(x_{1}-x_{2})+(x_{1}-x_{3})](x_{3}-x_{1})
    \notag\\&\quad+\frac{x_{1}}{x_{2}}[(x_{2}-x_{1})+(x_{2}-x_{3})](x_{3}-x_{2})\Big]~.
    \end{align}

    We calculate $T_{13}$ and $T_{23}$ similarly.
    \begin{align}
    T_{13}&=\frac{3x_{1}-1}{x_{1}}\dot{x}_{3}-\frac{3x_{3}-1}{x_{3}}\dot{x}_{1}
    \notag\\&=\frac{3x_{1}-1}{x_{1}}(1-\alpha x_{1}-\beta x_{2}- x_{3})x_{3}
    \notag\\&\quad-\frac{3x_{3}-1}{x_{3}}(1-x_{1}-\alpha x_{2}-\beta x_{3})x_{1}
    \notag\\&=\frac{\beta-\alpha}{2}\Big[\frac{x_{3}}{x_{1}}[(x_{1}-x_{2})+(x_{1}-x_{3})](x_{1}-x_{2})
    \notag\\&\quad+\frac{x_{1}}{x_{3}}[(x_{3}-x_{1})+(x_{3}-x_{2})](x_{2}-x_{3})\Big]~.
    \end{align}

    \begin{align}
    T_{23}&=\frac{3x_{2}-1}{x_{2}}\dot{x}_{3}-\frac{3x_{3}-1}{x_{3}}\dot{x}_{2}
    \notag\\&=\frac{3x_{2}-1}{x_{2}}(1-\alpha x_{1}-\beta x_{2}- x_{3})x_{3}
    \notag\\&\quad-\frac{3x_{3}-1}{x_{3}}(1-\beta x_{1}- x_{2}-\alpha x_{3})x_{2}
    \notag\\&=\frac{\beta-\alpha}{2}\Big[\frac{x_{3}}{x_{2}}[(x_{2}-x_{1})+(x_{2}-x_{3})](x_{1}-x_{2})
    \notag\\&\quad+\frac{x_{2}}{x_{3}}[(x_{3}-x_{1})+(x_{3}-x_{2})](x_{3}-x_{1})\Big]~.
    \end{align}

    Thus we calculate out each element of matrix $T$ on the plane $x_{1}+x_{2}+x_{3}=1$.

\item
    When $\gamma\neq0$:

    As
     \begin{align}
    \left\{
     \begin{array}{l}
        \frac{\partial\phi}{\partial x_{1}}=\gamma\frac{P}{O^{4}}[x_{2}x_{3}(x_{1}+x_{2}+x_{3})-3x_{1}x_{2}x_{3}] \\
        \frac{\partial\phi}{\partial x_{2}}=\gamma\frac{P}{O^{4}}[x_{1}x_{3}(x_{1}+x_{2}+x_{3})-3x_{1}x_{2}x_{3}] \\
        \frac{\partial\phi}{\partial x_{3}}=\gamma\frac{P}{O^{4}}[x_{1}x_{2}(x_{1}+x_{2}+x_{3})-3x_{1}x_{2}x_{3}]
      \end{array}
    \right.~,
    \end{align}
    \begin{align}
    S&=-\frac{\nabla
    \phi\cdot\mathbf{f}}{\mathbf{f}\cdot\mathbf{f}}I
    \notag\\&=\frac{\frac{\gamma^{2}P}{2O^{4}}\Big[(x_{1}-x_{2})^{2}+(x_{2}-x_{3})^{2}+(x_{3}-x_{2})^{2}\Big]}{\dot{x}_{1}^{2}+\dot{x}_{2}^{2}+\dot{x}_{3}^{2}}I~.
    \end{align}
    Since on the plane $x_{1}+x_{2}+x_{3}=1$, $S$ is not zero matrix except on the limit set ${(x_{1},x_{2},x_{3})|P=0, O=1}$. Thus on the plane the system is dissipative when $\gamma\neq0$.

    As for $T$:
    \begin{align}
    T&=-\frac{\nabla \phi\times \mathbf{f}}{\mathbf{f}\cdot\mathbf{f}}
    \notag\\&=-\frac{\gamma\frac{P^{2}}{O^{4}}\Big(\frac{O-3x_{i}}{x_{i}}\dot{x}_{j}-\frac{O-3x_{j}}{x_{j}}\dot{x}_{i}\Big)_{3\times3}}{\dot{x}_{1}^{2}+\dot{x}_{2}^{2}+\dot{x}_{3}^{2}}~.
    \end{align}
    Again, as all the trajectory will converge to the plane $O=x_{1}+x_{2}+x_{3}=1$, we just calculate the elements of above $3\times3$ matrix on the plane:
    \begin{align}
    \widetilde{T}_{ij}=\frac{1-3x_{i}}{x_{i}}\dot{x}_{j}-\frac{1-3x_{j}}{x_{j}}\dot{x}_{i}.
    \end{align}
    Here we use $\widetilde{T}_{ij}$ to denote the matrix elements so that they can be distinguished with the matrix elements $T_{ij}$ in the case of $\gamma=0$. Since $1-\alpha=\frac{\beta-\alpha-\gamma}{2}$ and $1-\beta=-\frac{\beta-\alpha+\gamma}{2}$ in the case of $\gamma\neq0$, we get $\widetilde{T}_{12}$, $\widetilde{T}_{13}$ and $\widetilde{T}_{23}$ with similar calculation:
    \begin{align}
    \widetilde{T}_{12}&=\frac{x_{1}}{x_{2}}[(x_{2}-x_{1})+(x_{2}-x_{3})][(1-\alpha)x_{2}+(1-\beta)x_{3}]
    \notag\\&\quad-\frac{x_{2}}{x_{1}}[(x_{1}-x_{2})+(x_{1}-x_{3})][(1-\alpha)x_{3}+(1-\beta)x_{1}]~.
    \end{align}
    We calculate $\widetilde{T}_{13}$ and $\widetilde{T}_{23}$ similarly.
    \begin{align}
    \widetilde{T}_{13}&=\frac{x_{1}}{x_{3}}[(x_{3}-x_{1})+(x_{3}-x_{2})][(1-\alpha)x_{2}+(1-\beta)x_{3}]
    \notag\\&\quad-\frac{x_{3}}{x_{1}}[(x_{1}-x_{2})+(x_{1}-x_{3})][(1-\alpha)x_{1}+(1-\beta)x_{2}]~.
    \end{align}
    \begin{align}
    \widetilde{T}_{23}&=\frac{x_{2}}{x_{3}}[(x_{3}-x_{1})+(x_{3}-x_{2})][(1-\alpha)x_{3}+(1-\beta)x_{1}]
    \notag\\&\quad-\frac{x_{3}}{x_{2}}[(x_{2}-x_{1})+(x_{2}-x_{3})][(1-\alpha)x_{1}+(1-\beta)x_{2}]~.
    \end{align}
    Thus we calculate out each element of matrix $T$ on the plane $x_{1}+x_{2}+x_{3}=1$.
\end{enumerate}

\bibliography{bib}

\end{document}